\documentclass[lettersize,journal]{IEEEtran}
\usepackage{amsmath,amsfonts}
\usepackage{algorithmic}
\usepackage{algorithm}
\usepackage{array}
\usepackage[caption=false,font=normalsize,labelfont=sf,textfont=sf]{subfig}
\usepackage{graphicx}
\usepackage{float}
\usepackage{subfig}
\usepackage{textcomp}
\usepackage{stfloats}
\usepackage{url}
\usepackage{verbatim}
\usepackage{graphicx}
\usepackage{cite}
\usepackage{color}
\newcolumntype{L}[1]{>{\raggedright\arraybackslash}m{#1}}
\newcolumntype{C}[1]{>{\centering\arraybackslash}m{#1}}

\begin{document}

\title{Mobile Communications in Intelligent Rail Transit: From LCX to PASS}

\author{Yiran Guo,~\IEEEmembership{Student Member,~IEEE,} Wei Chen,~\IEEEmembership{Senior Member,~IEEE,} Cong Yu,  \\Bo Ai,~\IEEEmembership{Fellow,~IEEE,} Yuanwei Liu,~\IEEEmembership{Fellow,~IEEE,} and Michail Matthaiou,~\IEEEmembership{Fellow,~IEEE}

\thanks{Yiran Guo, Wei Chen, Cong Yu, and Bo Ai are with the State Key Laboratory of Advanced Rail Autonomous Operation, and the School of Electronic and Information Engineering, Beijing Jiaotong University, China (e-mail:\{yiranguo, weich, cong\_yu, boai\}@bjtu.edu.cn).

Yuanwei Liu is with the Department of Electrical and Electronic Engineering, The University of Hong Kong, Hong Kong (e-mail: yuanwei@hku.hk).

Michail Matthaiou is with the Centre for Wireless Innovation (CWI), Queen’s University Belfast, U.K. (e-mail: m.matthaiou@qub.ac.uk).}
}

\maketitle

\begin{abstract}
Wireless communications in intelligent rail transit face harsh propagation conditions, including severe penetration loss, frequent blockages, and amplified large-scale fading. Existing leaky coaxial cables (LCX) provide wired-to-wireless conversion and stable coverage, but can be energy- and spectrum-inefficient, particularly at high carrier frequencies. Motivated by the growing demand for high-capacity and high-reliability rail services, this article introduces pinching-antenna systems (PASS), which are flexible waveguide-based architectures that enable reconfigurable radiation points with low deployment overhead and a natural fit to predominantly straight track geometries. We discuss the key benefits and deployment flexibility of PASS, evaluate their performance relative to LCX via representative simulations, and present a deep learning (DL)-enabled channel-estimation framework to cope with mobility-induced channel dynamics. Finally, we summarize the major open challenges for practical deployment and outline promising research directions.

\end{abstract}


\section{Introduction}
\label{intro}
\begin{figure*}[!t]
\centering
\includegraphics[width=7in]{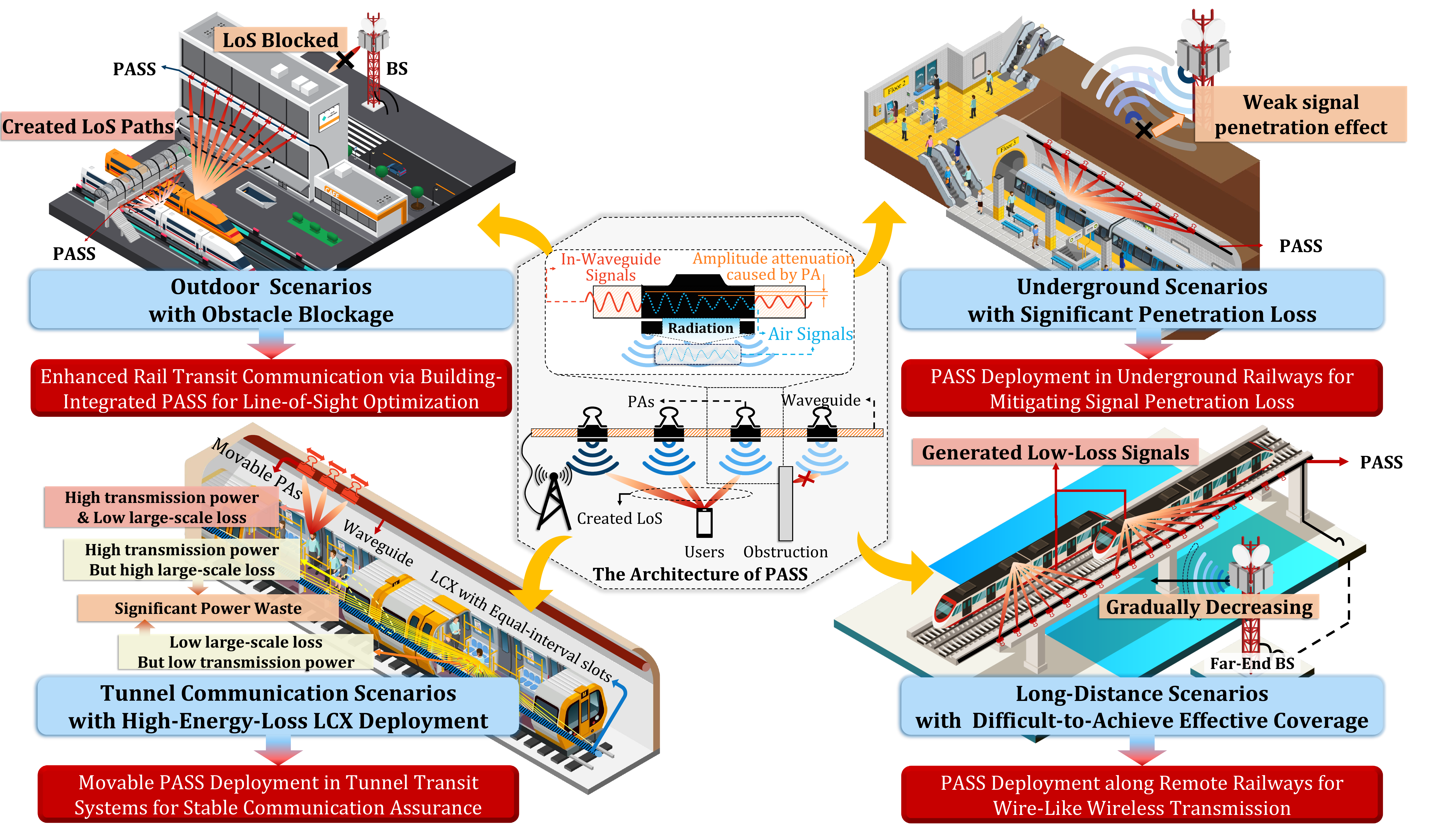}
\caption{Typical intelligent rail transit scenarios and examples of PASS deployments.}
\label{scenario}
\end{figure*}

Rail transit is a cornerstone of modern transportation thanks to its high efficiency and low environmental footprint. Yet, as rail systems evolve toward intelligence and autonomy, existing mobile communication infrastructures are struggling to satisfy the rapidly growing demands of rail services. Following the visions and frameworks of the International Telecommunication Union (ITU) and the International Union of Railways (UIC), next-generation intelligent rail transit requires robust, ubiquitous connectivity to enable applications, such as automatic train operation, fully autonomous driving, and high-rate infotainment data delivery \cite{NG_IRT}. Meeting these heterogeneous requirements calls for communication solutions that simultaneously enhance reliability, improve energy efficiency, and increase data rates. In this article, we consider next-generation intelligent rail transit scenarios incorporating a novel antenna technology to increase system capacity, improve energy efficiency, and provide high-quality connectivity.

In practice, rail transit communications are constrained by harsh and highly heterogeneous propagation environments. In urban corridors, high-rise buildings and terrain block the direct link between base stations (BSs) and fast-moving trains; while scattered and refracted components may provide intermittent coverage, the absence of a line-of-sight (LoS) path can cause an amplified degradation in link quality. In underground sections, severe penetration loss further hinders reliable signal acquisition and data transmission. In remote railways, where track-side BS deployment is sometimes impractical, trains may rely on distant BSs and can maintain LoS connectivity only under favorable unobstructed conditions. To cope with these challenges, hybrid wired--wireless solutions based on leaky coaxial cables (LCX) tunnel deployments have been adopted, which radiate signals into free space via periodically slotted outer conductors \cite{LCX_JSAC}. Although LCX can provide stable coverage, their fixed physical architecture leads to inefficient energy and spectrum utilization as trains move, whereas the associated installation cost makes multi-stream transmission difficult to realize.

Pinching-antenna systems (PASS) \cite{PASS_DOCOM} have recently attracted considerable attention as a flexible antenna architecture capable of creating controllable radiation points and, in turn, establishing near-direct links in challenging environments. For comparison, LCX realize wired-to-wireless conversion via periodically slotted outer conductors. The slot spacing, the relative permittivity of the dielectric layer, and the signal wavelength jointly determine the peak radiation angle \cite{LCX_Receiver} as well as the amplitude attenuation and phase evolution along the cable. By contrast, the pinching-antenna architecture developed and field-validated by NTT DOCOMO in 2021 uses low-loss dielectric waveguides as the transmission medium \cite{PASS_DOCOM}, which can significantly reduce transmission loss at high frequencies compared with conventional coaxial cables. More importantly, PASS enable flexible and controllable radiation points along the waveguide. These points are created by dielectric particles extruded onto the waveguide surface, acting as clamp-like elements (Fig.~\ref{scenario}), termed pinching antennas (PAs). The coupling length and the PA--waveguide spacing govern the attenuation and phase shift of the radiated signals, enabling spatial degrees of freedom that extend well beyond a single wavelength.

Unlike the fixed slot locations in LCX, PAs in PASS can be repositioned along the dielectric waveguide and selectively activated to adapt the radiating aperture to user mobility and service demands \cite{Movable_PA1,Movable_PA2,Activate_PA}. If a PA is deactivated (i.e., not clamped to the waveguide), it does not radiate and the guided energy continues to propagate toward downstream PAs, enabling dynamic channel reconfiguration with minimal overhead. This flexibility has motivated a growing body of optimization work, including continuous-position PA deployment for improved sensing robustness \cite{Movable_PA1}, jointly optimization of PA placement/activation to meet quality-of-service constraints \cite{Movable_PA2}, and discrete PA activation strategies for sum-rate maximization in non-orthogonal multiple access (NOMA) systems \cite{Activate_PA}. For clarity, Table~\ref{tab:lcx_pass} summarizes key distinctions between LCX variants and PASS configurations. Overall, PASS offer a cost-effective, reconfigurable radiating structure, where PA positions and activation states can be controlled (e.g., via solid state relays (SSRs)) so that radiation occurs only at selected locations.

Notably, the predominantly straight-line track geometry in intelligent rail transit provides a favorable substrate for PASS deployment, making them a practical candidate for real-world rail communication systems. As illustrated in Fig.~\ref{scenario}, PASS can be mounted on roadside buildings or deployed close to the track in enclosed sections to bypass blockages and create alternative LoS paths. For viaducts and elevated railways, installing PASS along parapets or beneath the deck enables geometry-matched coverage along the track direction. In underground sections, the BS can feed PASS through a cable connection so that signals are radiated toward trains and track-side equipment, thereby alleviating the penetration-loss bottleneck. For long-distance railways, PASS can be laid along the track to form wired links that act in a relay-like manner, reducing free-space path loss over extended spans. In tunnels, replacing LCX with PASS equipped with movable or selectively activated PAs further enables a ``chasing'' mode, where the active PA follows the train to maintain the nearest high-quality link.
By continually maintaining the nearest direct path, this mechanism strengthens the received power while reducing propagation loss and unnecessary radiation, leading to more stable and energy-efficient connectivity for intelligent rail transit.

In this article, we first compare the LCX solution, which has been widely deployed in practical tunnel environments, with PASS, a promising but not yet implemented in the railway transit technology. We briefly introduce the architectures and implementation principles of both LCX and PASS, and analyze their theoretical performance upper bounds in typical tunnel scenarios. Subsequently, we highlight the advantages of PASS over conventional antenna architectures when applied to intelligent rail transit systems and conduct simulation experiments to evaluate their performance gains in energy saving. In Section~\ref{Section_DL_Power}, we discuss the challenges that PASS will face in time-varying rail transit communication channels and explore the use of deep learning (DL)-based methods to enhance communication performance of intelligent rail transit systems with PASS. Finally, we summarize and analyze the potential difficulties and challenges associated with the practical deployment of PASS in intelligent rail transit.

\section{Evolution from LCX to PASS for Intelligent Rail Transit}
\label{Section_2}
\begin{table*}[!t]
\centering
\caption{Comparative Overview of LCX Variants and PASS Configurations.}
\label{tab:lcx_pass}
\renewcommand{\arraystretch}{1.45}
\setlength{\tabcolsep}{3.5pt}
\begin{tabular}{
  L{2.3cm}
  L{1.2cm}
  L{3.2cm}
  L{3cm}
  L{3cm}
  L{3.4cm}
}
\hline\hline
{\textbf{System}} &
{\textbf{Ref}} &
{\textbf{Antenna Reconfig}} &
{\textbf{Deployment Cost}}&
{\textbf{Spectral Efficiency}} &
{\textbf{Energy Efficiency}}
\\
\hline\hline
 
Single-Ended LCX &
\cite{LCX_ta1}&
Fixed&
Low &
Low; Worsens at cable end &
Low; High coupling loss accumulates \\
\hline
 
Double-Ended LCX &
\cite{LCX_Receiver}&
Fixed&
High; Dual-feed hardware&
Moderate; Improved over single-ended &
Moderate \\
\hline
 
Segmented LCX &
\cite{LCX_Interval}&
Fixed&
Highest; Active repeaters \& Sync hardware &
High; Best among LCX variants &
Moderate; Repeater power overhead \\
\hline\hline
 
PASS Fix  &
\cite{PASS_DOCOM}&
Detachable &
Low &
Moderate; Fixed PA positions&
Moderate; Short free-space path to nearest PA  \\
\hline
 
Active PASS &
\cite{PASS_CE,Ding_CE}&
Detachable; Electronically switched &
Moderate; SSR switching logic required&
High; Near-ideal LoS link maintained &
High; Energy concentrated on user \\
\hline
 
Movable PASS &
\cite{Movable_PA2,Movable_PA1}&
Detachable; Mechanical actuators &
High; mechanical tracking&
Highest; Theoretical upper bound &
Highest; Minimum path loss at all times 
\\
\hline\hline
\end{tabular}
\end{table*}


With a fixed total radiated power, LCX release energy through a sequence of slots while the remaining guided power gradually diminishes along the cable. Hence, near the cable end, the closest slot benefits from relatively small large-scale fading due to short separation but radiates limited power; conversely, slots closer to the feed point radiate more power yet suffer much stronger path loss because of the longer distance. This inherent power--loss mismatch allows LCX to provide good link quality near the feed point, but leads to pronounced performance degradation toward the far end of the cable. To alleviate this longitudinal attenuation, evolutionary LCX architectures, such as double-ended and segmented LCX, have been proposed. However, these approaches typically rely on specific physical slotting designs or spatial segmentation strategies that configure independent radio frequency (RF) repeaters for each physical section. Such active relay-based enhancements compromise system simplicity and may introduce inter-section interference, amplify complexity, and stringent synchronization overhead, while substantially increasing engineering cost. In ultra-long confined spaces, the incorporation of repeaters, duplex amplifiers, and long-distance feed networks can significantly erode the cost-effectiveness of conventional cabling solutions.
In contrast, deploying PASS in rail transit communication systems can effectively mitigate, or even eliminate, this issue via pre-designed PAs. In this section, in addition to the LCX deployment case, we introduce three PA configurations as follows:




\begin{itemize}
    \item \textbf{Case~1 (Equidistant PAs):}
    Multiple PAs are uniformly deployed along the waveguide and all are activated. By tailoring the PA length and coupling strength (or equivalently the effective PA density), the radiated power can be balanced across PAs without over-populating the waveguide input end. Consequently, users near the waveguide end can still be served by sufficiently strong terminal radiation, while the short distance to the nearest terminal PAs helps sustain a stable average channel capacity along the trajectory.

    \item \textbf{Case~2 (Movable PAs):}
    This configuration represents an ideal upper bound: a PA continuously tracks the user so that the dominant radiation point remains closest to the user, thereby minimizing path loss while focusing the radiated energy. Nevertheless, for high-mobility rail scenarios, mechanically repositioning PAs fast enough for real-time tracking is generally impractical.

    \item \textbf{Case~3 (Partially Activated PAs):}
    PAs are deployed at equal intervals as in Case~1, whereas only the PAs nearest to the user are activated. This strategy can be viewed as a discrete, practically implementable approximation of Case~2. With low-latency SSRs, the activation state of PAs can dynamically adapt to user movement, effectively ``following'' the user to maintain a stable, low-loss LoS path while avoiding unnecessary radiation in other regions.With low-latency SSRs, the activated PA can switch to follow the user, maintaining a stable low-loss LoS path while avoiding unnecessary radiation elsewhere.
\end{itemize}

\begin{figure}[!t]
\centering
\includegraphics[width=3.4in]{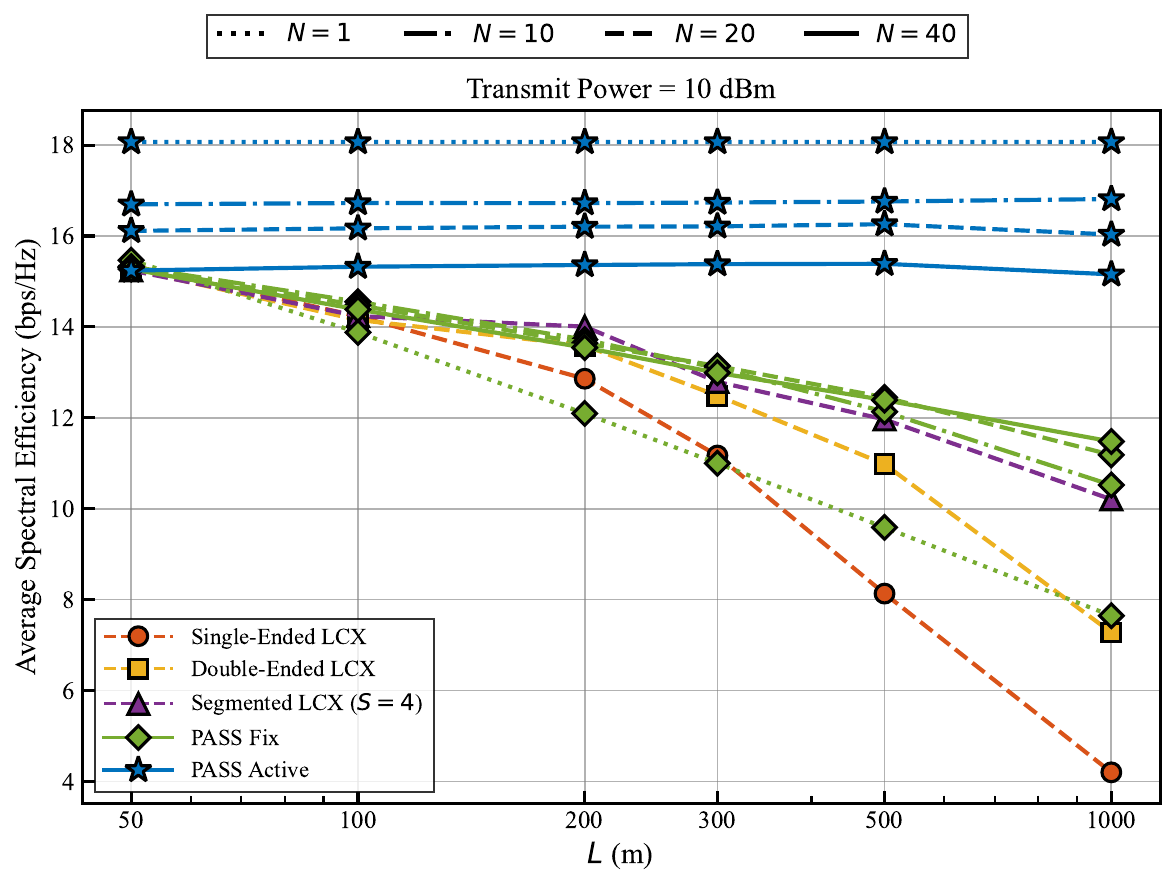}
\caption{Comparison of the maximum achievable spectral efficiency with LCX and PASS in intelligent rail transit systems.
 In the ``PASS Fix'', $N$ equally spaced PAs are deployed. In the ``PASS Active'', with PA spacing exceeding half a wavelength, PAs are deployed with 1.2195~m intervals, and the maximum number of activated PAs is set to $N$.}
\label{EX1_LCX_vs_PASS}
\end{figure}

The effectiveness of PASS is now evaluated through simulation experiments, with results presented in Fig.~\ref{EX1_LCX_vs_PASS}. Five use-case are considered: single-ended LCX, double-ended LCX, segmented LCX with $S=4$ segments, Case~1 (denoted as ``PASS Fix''), and Case~3 (denoted as ``PASS Active'') in a straight line environment with length $L$ ($x$-axis), width $D_y = 20$~m ($y$-axis), and height $D_z = 5$~m ($z$-axis). The average spectral efficiency is depicted for each scenario along the entire straight line path. Both LCX and PASS are deployed along the $x$-axis, starting at coordinates $[0, D_y/2, D_z]$ and ending at $[L, D_y/2, D_z]$, which also defines the user’s trajectory with the $z$-coordinate set to 0. A total of 10,000 data points are sampled along the x-axis; the spectral efficiency was calculated at each point, and the mean value is computed to evaluate the average spectral efficiency along this path. Noise power is fixed at $-120$~dBm and the signal frequency at 28~GHz. For both LCX and PASS, signal propagation in free space follows the same free-space loss model, varying only with distance. For LCX, the slot interval is set to 0.08~m, the amplitude attenuation factor to 9.8~dB/100~m \cite{LCX_JSAC}, and the relative permittivity to 1.26. For PASS, the relative power parameter is 0.95 and the effective refractive index is 1.4. The track length $L$ is selected from $L \in\{50, 100, 200, 300, 500, 1000\}$~m. The total transmit power is set to 10~dBm. For dual ended LCX and segmented LCX configurations, the total transmission power $P_t$ is equally divided into $P_t/2$ and $P_t/4$, respectively, to ensure a fair comparison.

From the experimental results, we observe that PASS almost consistently outperform LCX, except for the ``PASS Fix'' configuration when $N=1$ and some specific $L$. For example, at $N=1$, the ``PASS Active'' case provides a gain of approximately $2.71\,\mathrm{bps/Hz}$ over single-end LCX at $L=50\,\mathrm{m}$ and $13.87\,\mathrm{bps/Hz}$ at $L=1000\,\mathrm{m}$. At the same time, the results indicate that dual-ended LCX and segmented LCX achieve higher channel capacity than single-ended LCX, particularly in long line scenarios. When $L = 1,000$ m, the performance gap between dual-ended LCX and movable PA under the condition of $N = 1$ decreases to 10.79 bps/Hz, while the corresponding gap for segmented LCX is reduced to 7.86 bps/Hz. Nevertheless, compared with LCX type antennas, PASS still provide performance gains, even when the PA positions are fixed, especially in long line scenarios.

\section{Energy-Efficient and Flexible Railway Wireless Communication Enabled by PASS}
\label{sec:general_for_railway}

In this section, in addition to the enhanced spectral efficiency, we will discuss three additional promising benefits that PASS could bring to intelligent rail transit wireless communication systems.

\subsection{Railway Track Geometry Suitability}
\label{sub:flexible}
The low-complexity structure of PASS enables straightforward deployment along railway infrastructure, including tunnel ceilings, platform roofs, trackside barriers, and viaduct structures as shown in Section~\ref{scenario}. In particular, PASS form LoS-dominant, short-range wireless links by extensively deploying waveguides and optimally activating PAs located near the trains.  

\par
PASS introduce additional degrees of freedom for beamforming optimization through dynamic PA repositioning along waveguides. In continuous activation mode, PAs mounted on motorized slides can adjust their positions in real-time, enabling pinching beamforming by tuning PA locations to achieve constructive interference at receivers while mitigating free-space path loss. This architecture also supports parallel multi-waveguide configurations fed by independent RF chains, transforming the infrastructure into a flexible distributed antenna array that creates dedicated high-quality LoS propagation environments for individual users. By coordinating the optimal activation and precise positioning of PAs across these independent waveguide links, the system can synthesize beams tailored to specific user locations \cite{PASS_Beamforming2}. This strategy can effectively establish exclusive spatial channels for each receiver, isolating them from interference and mitigating path loss. Consequently, PASS can simultaneously serve multiple users with parallel data streams, significantly enhancing spectral efficiency and network capacity through near-field spatial multiplexing. 

In addition, PAs can be added or removed to accommodate time-varying traffic demands. For example, when a metro line experiences sustained ridership growth, the operator can densify the PA deployment on existing waveguides instead of rebuilding the infrastructure. Although evolved LCX architectures (e.g., double-ended and segmented LCX) provide limited design-time flexibility via slot-density grading and spatial segmentation, their radiation behavior is ultimately dictated by fixed slot patterns and the dielectric properties of the insulating layer; thus, any post-deployment reconfiguration typically requires physical replacement. In PASS, by contrast, radiation points are defined by the clamping locations of PAs on the dielectric waveguide, and their activation states can be configured on demand. This distinction allows PASS to preserve spatial reconfigurability throughout its operational lifetime, adapting to train motion and traffic fluctuations without modifying the underlying waveguide infrastructure.

\subsection{Handover Reduction}
\label{sub:handover}
Beyond flexible coverage, PASS have the potential to fundamentally transform mobility management in railway wireless systems. The extended reach of waveguide-based transmission drastically reduces handover frequency compared to cellular networks' limited cell dimensions, directly enhancing reliability for safety-critical railway operations. PASS fundamentally alleviats handover pressure through an extended waveguide reach. In contrast to existing cells limited by propagation physics to sub-kilometer ranges in urban railway environments, a single PA waveguide can span significantly longer distances with negligible signal degradation. As trains traverse these extended coverage zones, PAs dynamically activate and deactivate based on train location tracked through positioning systemsand train management systems, but crucially, no handover occurs, as the same RF chain and baseband processing maintain the connection throughout the waveguide span. This architectural advantage translates to dramatic handover reduction across typical railway corridors. For intercity routes, PASS can reduce handover events by over an order of magnitude compared to cellular systems. Every handover that is avoided removes a possible point of failure, which is especially important for preventing service interruptions. The reduced signaling overhead from fewer handovers also results in increasing system throughput.

\subsection{Significant Energy Savings}
PASS achieve high energy efficiency by separating long-distance signal transport (via waveguides) from short-range radiation (via PAs), allowing each function to be optimized independently and avoiding the compromises inherent in conventional systems.

\textbf{Waveguide Transport Efficiency:} In contrast to free-space transmission, where path loss grows rapidly with distance, dielectric waveguides confine electromagnetic energy and transport it over long spans with very low attenuation, thereby reducing (or even avoiding) the periodic amplification that dominates the power budget of legacy trackside deployments. Although LCX also guide signals along the railway, its coupling loss accumulates continuously along the cable and cannot be adaptively offset after installation \cite{LCX_ta1}. Dielectric waveguides circumvent this limitation, enabling near-lossless signal delivery so that the power budget can be devoted primarily to the short-range wireless radiation link rather than to compensating for infrastructure-induced loss.

\textbf{Last-Meter Radiation Optimization:} In conventional cellular deployments, BSs must illuminate large areas under highly variable propagation conditions (e.g., blockage and scattering from buildings and terrain), which typically forces conservative transmit-power margins. In PASS, radiation is confined to the last-meter link between the nearest active PA and the train, where the geometry is predominantly LoS and the path loss is well approximated by a free-space model. This predictability enables fine-grained power control and directional radiation toward the instantaneous train location, achieving the target received signal-to-noise ratio (SNR) with substantially lower radiated power than omnidirectional trackside transmission \cite{Pass_power}. Moreover, since train trajectories and timetables are known a priori, PA activation can be scheduled deterministically: only a small set of PAs around the train are turned on while the remaining PAs stay in a low-power sleep state, thereby reducing the average energy consumption compared with always-on cellular BSs.

\begin{figure}[!t]
\centering
\includegraphics[width=3.4in]{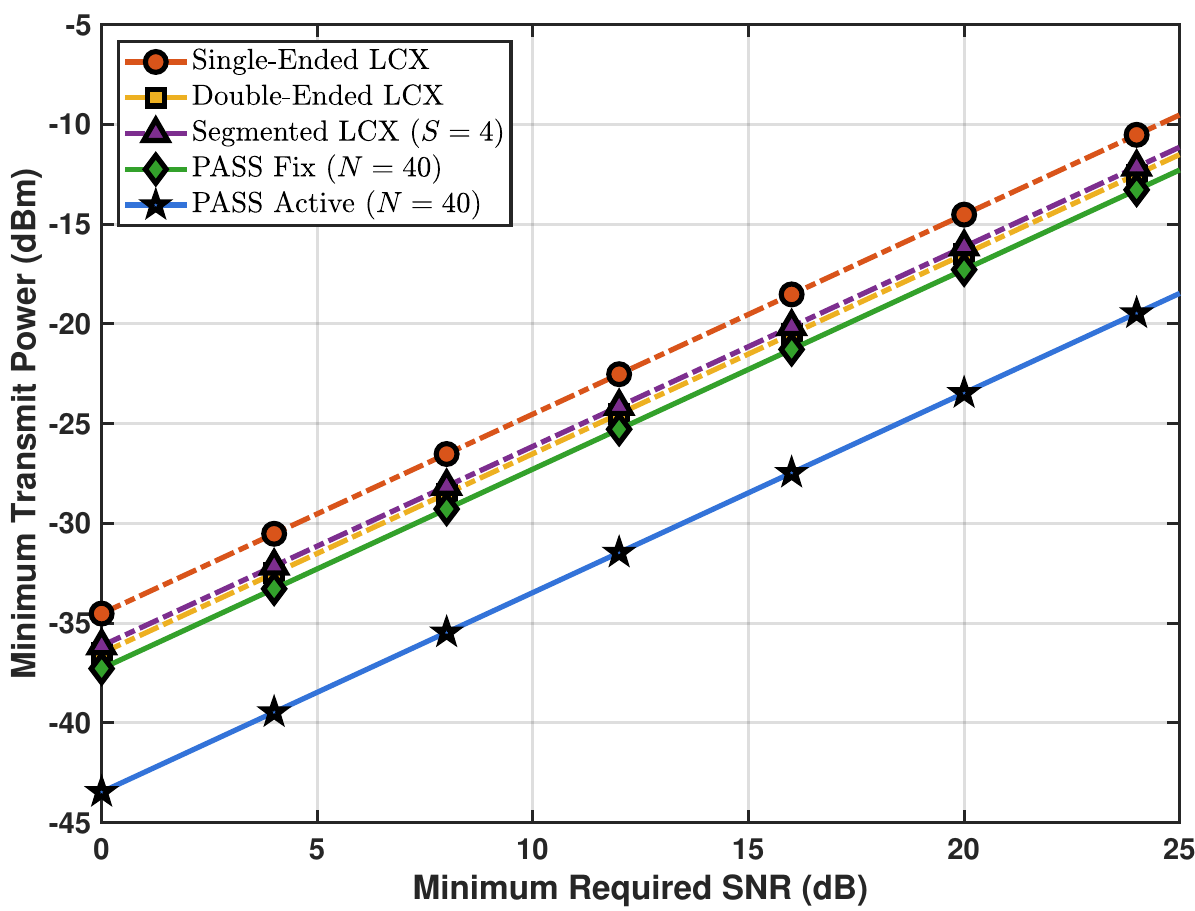}
\caption{Transmit power versus the users' minimum SNR.}
\label{fig:sectionIII_power}
\end{figure}

The energy efficiency of PASS is evaluated by comparing the minimum transmit power required to meet a target SNR, with results summarized in Fig.~\ref{fig:sectionIII_power}. We consider five representative architectures: single-ended LCX, double-ended LCX, segmented LCX with $S=4$ segments, PASS Fix with $N=40$ PAs, and PASS Active with $N=40$ PAs over a $40$~m active span. The simulation geometry and the electrical parameters of the LCX configurations follow the setup in Section~\ref{Section_2}. For each target SNR threshold in $[0,25]$~dB, we compute the minimum transmit power that guarantees the required SNR at the user. Single-ended LCX exhibits the largest power demand because the guided signal experiences cumulative attenuation along the entire cable, leading to weak coverage near the far end. Double-ended and segmented LCX mitigate this effect by reducing the maximum attenuation distance, and therefore achieve similar power requirements. In contrast, both PASS variants consistently require less transmit power since the dense PA deployment bounds the user--radiator distance. PASS Active attains the lowest power by selecting the nearest PA for radiation, yielding an approximately $10$~dB reduction relative to single-ended LCX. Overall, the results validate PASS as more energy-efficient coverage solutions for confined linear railway environments, with dynamic PA activation providing the most pronounced gain.

\section{DL–Empowered Communication Enhancement for PASS}
\label{Section_DL_Power}
User mobility in intelligent rail transit introduces stringent latency and robustness requirements that challenge the practical operation of PASS. In particular, accurate channel state information (CSI) for the links between the user and a large number of PAs is a prerequisite for PA selection, activation control, and coherent transmission. However, PASS channel acquisition is fundamentally complicated by the intertwined free-space and waveguide propagation mechanisms: the observed pilots are jointly shaped by the over-the-air PA--user channels and the in-waveguide coupling/transport responses, yielding an ill-conditioned and often underdetermined estimation problem \cite{PASS_CE, Ding_CE}. While conventional schemes (e.g., least squares (LS) estimation with sequential PA activation) are applicable, their training latency scales roughly linearly with the number of PAs, which is incompatible with the short coherence time in high-mobility rail scenarios. Moreover, mobility-induced channel aging quickly invalidates previously estimated CSI, forcing frequent re-training and incurring substantial pilot overhead.

\begin{figure}[!t]
\centering
\includegraphics[width=2.5in]{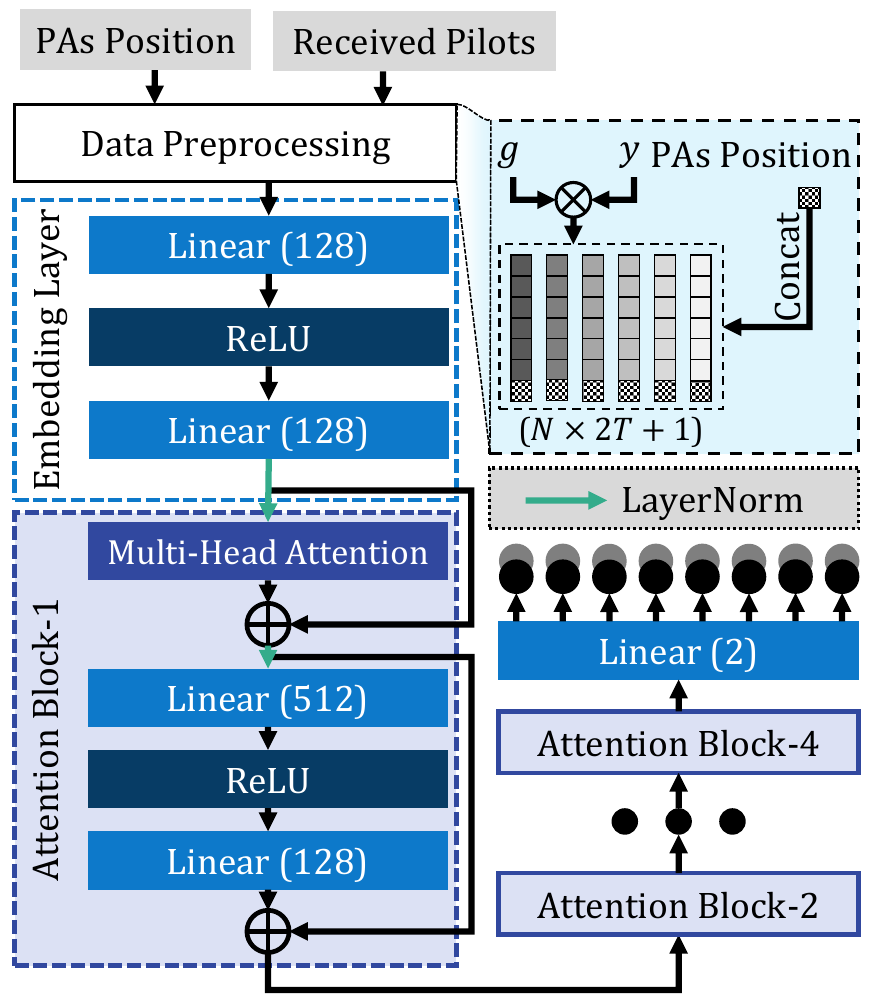}
\caption{PAformer network architecture based on multi-head attention, featuring eight heads, four attention blocks, and 128 hidden neurons.}
\label{PAformer_architecture}
\end{figure}

Recent advances in DL provide a promising toolset for addressing these challenges via high-dimensional inference, including channel estimation, CSI reconstruction, and antenna/channel extrapolation. PASS channels exhibit strong structural regularities: given the fixed track and the known PA layout, the dominant geometric relationship between the user and the PAs induces a position-dependent CSI manifold with pronounced spatial correlation. Such structure can be learned from data and exploited to map low-dimensional pilot observations to high-dimensional CSI, offering improved estimation accuracy under mobility while reducing training latency and pilot overhead. In the following, we adopt the PAformer network in Fig.~\ref{PAformer_architecture}, a Transformer-based CSI acquisition framework proposed in~\cite{PASS_CE}.

The PA deployment follows the ``Equidistant PAs'' strategy with 16 PAs. The scenario remained consistent with Section~\ref{Section_2}, where the user moves along a straight path with a speed of 60~km/h. The pilot interval, which also serves as the sampling interval, was set at 0.625~ms \cite{Pre_Interval}, while the initial communication location is randomly assigned along the track. The dataset is generated using MATLAB, where the training, validation, and test sets consist of 70{,}000, 30{,}000, and 10{,}000 samples, respectively. The initial learning rate is set to $10^{-3}$ and decays to a minimum of $10^{-6}$. When the validation loss fails to decrease for 20 consecutive epochs, the learning rate is reduced by half. The maximum number of training epochs is set to 300. The model is trained using data uniformly sampled over $\mathrm{SNR} \in [0 \text{ dB}, 20 \text{ dB}]$. The loss function is defined as the mean squared error (MSE) between the true channel and the estimated channel.

\begin{figure}[!t]
\centering
\includegraphics[width=3.4in]{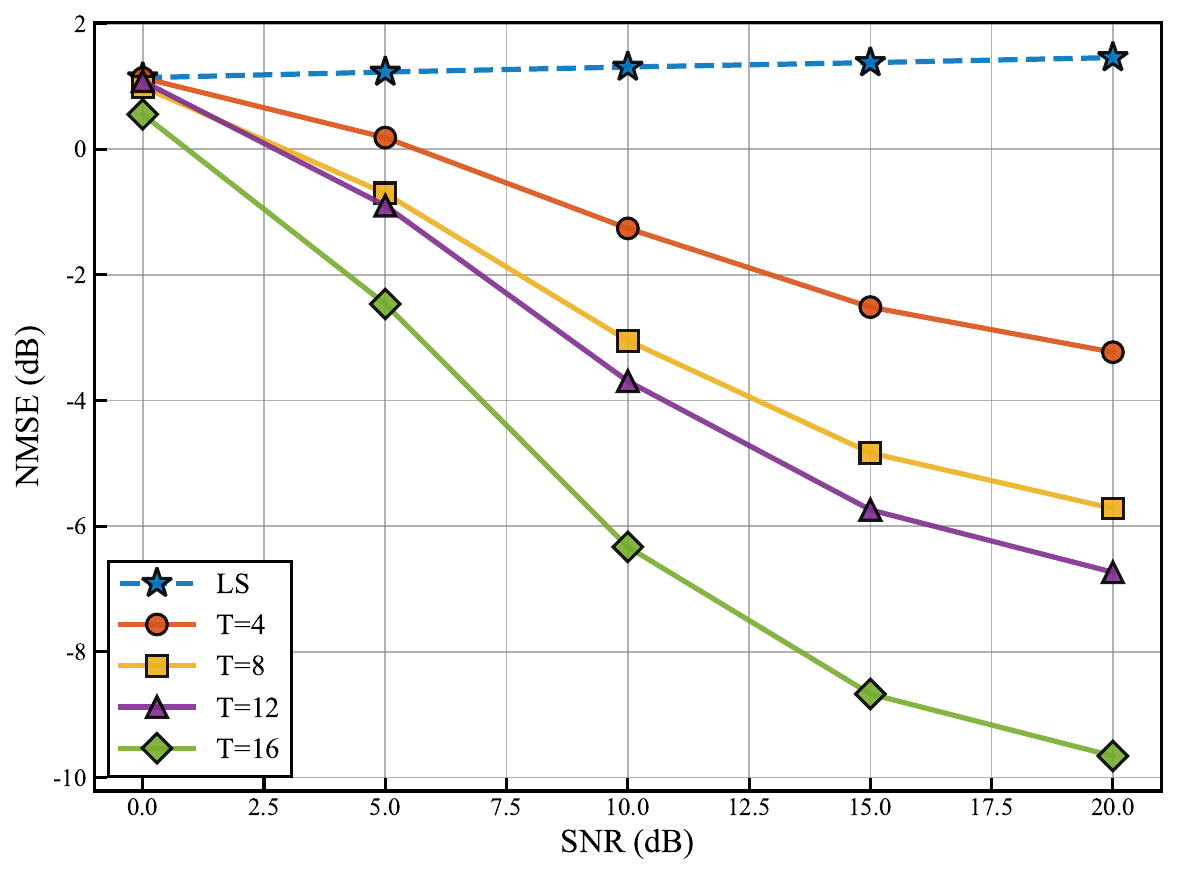}
\caption{Comparison of channel estimation accuracy between the PAformer network and LS algorithms under varying historical information lengths in mobile scenarios.}
\label{EX2_AI_CE}
\end{figure}

We compare the LS baseline with DL-based channel estimation under different historical pilot lengths $T$, with results reported in Fig.~\ref{EX2_AI_CE}. The DL-based method consistently outperforms LS in mobile scenarios. This is because LS primarily recovers the CSI of the currently activated PA at the most recent pilot instant, whereas the CSI associated with inactive PAs becomes outdated due to channel aging, which degrades the overall estimation quality when many PAs are involved. By contrast, the DL estimator leverages temporal context in the historical pilots and learns the underlying channel structure of PASS, leading to more reliable CSI reconstruction under mobility. For instance, at $\mathrm{SNR}=20$~dB, the proposed method achieves $-3.23$~dB with $T=4$, and further improves to $-9.66$~dB with $T=16$, corresponding to an 11.12~dB gain over LS.

The experimental results in Fig.~\ref{EX2_AI_CE} demonstrate the effectiveness of DL-enabled channel estimation in PASS. DL can be employed to deal with other challenges of PASS in mobile scenarios. For example, DL can be employed to predict future CSI and thereby reduce pilot overhead; perform user localization and trajectory tracking for timely PA switching, activation, or movement; and design end-to-end DL-based transmitters and receivers for PASS (applicable to pilot-overlay or pilot-free cases) to reduce waveguide deployment overhead while maintaining comparable system throughput.

\section{Opportunities and Challenges}
The largely straight railway track geometry naturally matches the linear physical structure of PASS deployments. With high flexibility and reconfigurable radiation/activation patterns, PASS represent a promising paradigm for improving spectral efficiency and user experience while enabling energy-efficient communications in intelligent rail transit systems. Nevertheless, PASS research is still in its infancy, and translating the concept into real-world deployments entails a number of nontrivial technical hurdles. In this section, we identify and analyze the key challenges that must be addressed for practical implementation.

\subsection{Coupling Relationships Among PAs}
Thanks to the intrinsically coupled waveguide-fed radiation, multiple PAs mounted on the same waveguide can emulate a distributed antenna array and generate multiple LoS components, which are beneficial for diversity and beamforming. However, when the PA--waveguide coupling length and the inter-PA spacing are fixed, the entire waveguide tends to behave as a single guided propagation medium, thereby fundamentally limiting the number of independently supportable spatial data streams. In practice, the coupling length is primarily determined by hardware/packaging constraints and is hard to adjust on the fly; likewise, rapidly reconfiguring the inter-PA spacing is infeasible in high-mobility rail scenarios. This "single-stream bottleneck" restricts the attainable multiplexing gain and, ultimately, the user capacity of PASS.

To overcome this limitation, existing work has explored integrating PASS with NOMA to serve multiple users \cite{Activate_PA}. However, NOMA relies on sufficiently heterogeneous user channels and power disparities; in dense environments, such as tunnels and underground railways, these conditions may not hold, limiting the number of simultaneously supportable users. Moreover, in single-user scenarios, NOMA does not increase the spatial degrees of freedom. A more direct approach is to deploy multiple parallel waveguides and interface them with multiple RF chains through a beamforming network, thereby enabling multi-stream transmission and improving both capacity and user support \cite{PASS_Beamforming,PASS_Beamforming2}.

A remaining open issue is the inter-waveguide (and inter-antenna) coupling in such multi-waveguide deployments. Prior studies on multi-LCX systems have shown that improper cable spacing can significantly degrade channel quality due to mutual coupling \cite{LCX_Interval}. Similar coupling-induced performance loss may arise in PASS, and its characterization and mitigation are essential for fully realizing the benefits of multi-waveguide PASS architectures.

\subsection{The Practical Implementation of PA Movement and Switching}
In Section~\ref{Section_2}, we identified the main contributors to the performance gains of PASS and quantified the spectral-efficiency advantages under different deployment cases. A key enabler is the ability to switch and activate PAs in a mobility-aware manner so that the effective serving aperture follows the user. However, realizing such dynamic operation in practice critically depends on the PA activation latency and the underlying hardware/firmware design, both of which remain insufficiently explored in the literature. In addition, the switching mechanism and the switching cadence must be jointly optimized with respect to mobility, channel dynamics, and control/signaling overhead, calling for systematic design guidelines and, eventually, unified deployment standards for PASS in intelligent rail transit systems.

The ``Partially Activated PAs'' use-case can be viewed as a pragmatic, suboptimal surrogate for fully ``Movable PAs''. To unlock the full potential of PASS, one direction is to develop PAs (or PA modules) that can physically track high-speed users with sufficiently fine spatial resolution. As a more implementable alternative, a hybrid architecture could deploy discrete PAs while permitting each PA limited local mobility (e.g., within a small mechanical range). With predictive localization and trajectory tracking, such PAs could be pre-positioned to approximate continuous spatial following, thereby narrowing the gap to the theoretical capacity upper bound of PASS in intelligent rail transit environments.

\subsection{Implementation of Uplink Communication}
Ensuring symmetric uplink and downlink performance is essential for end-to-end deployment. While PASS have been experimentally applied for downlink transmission \cite{PASS_DOCOM}, their uplink operation still lacks rigorous theoretical analysis and measurement-based verification, since the widely assumed channel reciprocity may not hold for waveguide--PA coupling. Several fundamental issues remain open: (i) whether the electromagnetic field distributions are sufficiently consistent between signal injection into the waveguide and leakage radiation into free space; (ii) whether the propagation direction and modal content of the injected signal can be reliably controlled inside the waveguide; and (iii) whether the PA-induced attenuation and phase response are reciprocal between transmit and receive modes.

A practical workaround is to employ LCX as auxiliary receive antennas \cite{Movable_PA1}, leveraging their well-established bidirectional transmit/receive capability \cite{LCX_Receiver}. However, adding a parallel LCX system would increase the deployment cost and infrastructure complexity. Therefore, dedicated uplink receiver designs tailored to PASS---including coupling structures, low-noise front ends, and calibration procedures---remain an important research direction. Alternatively, integrated LCX--PASS transceiver architectures are promising, where LCX will primarily support uplink reception and PASS will serve as the main downlink radiator, enabling cooperative multi-waveguide joint communications that exploits the complementary strengths of both media.

\section{Conclusion}
In this article, we introduced PASS as flexible antenna architectures for intelligent rail transit communications, with the goal of improving spectral efficiency, achievable data rates, and energy efficiency. We first reviewed the fundamental concept of PASS and discussed why their structural adaptability is well suited to rail scenarios. We then compared PASS with conventional LCX systems and validated the performance gains of PASS through simulations. Next, we highlighted key deployment benefits, including flexible installation, reduced handover frequency, and improved energy efficiency. To cope with mobility-induced channel dynamics, we further presented a DL-enabled design paradigm and demonstrated its effectiveness for PASS channel estimation. Finally, we discussed the major open challenges for real-world deployment and outlined potential solution directions.
\bibliographystyle{IEEEtran}
\bibliography{myref}

\end{document}